\documentstyle[a4wide,epsf,11pt,titlepage]{article}

\newcounter{nref}
\newcommand{\bbib}{%
  \renewcommand{\refname}{\large\bf References}%
  \begin{thebibliography}{99}%
  \setcounter{enumiv}{\thenref}}
\newcommand{\ebib}{%
  \end{thebibliography}%
  \setcounter{nref}{\arabic{enumiv}}}
\newcommand{\head}[3]{%
  \setcounter{nref}{0}%
  \thispagestyle{empty}%
  \section*{\LARGE\bf #1}%
  \stepcounter{section}%
  \addcontentsline{toc}{section}{#1}%
  \large\itshape%
  #2\\\vspace{0.1pt}\\%
  #3%
  \normalsize\upshape%
  \bigskip}

\begin{document}


\head{Global Anisotropies in Supernova Explosions\\
      and Pulsar Recoil}
     {
        L. Scheck$^1$,
        T. Plewa$^{2,3}$,
        K. Kifonidis$^1$,
        H.-Th. Janka$^1$,
        E. M{\"u}ller$^1$
     }{
	    $^1$ Max-Planck-Institut f\"ur Astrophysik, Garching, Germany     \\
        $^2$ Center for Astrophysical Thermonuclear Flashes, Chicago, USA \\
        $^3$ Nicolaus Copernicus Astronomical Center, Warsaw, Poland
     }

\vspace{0.5cm}

Observations show that neutron stars can have space velocities much
higher than those of their progenitors. On average pulsar birth
velocities are in the range of 200-500 km/s and significant fraction
might move with more than 1000 km/s (see \cite{scheck.ACC02} and
references therein).  The powerful and highly variable acceleration
mechanism that leads to the measured velocities is still unclear. It
is possible that a recoil velocity is imparted to the neutron star by
explosion asymmetries at the moment of its birth. Herant
\cite{scheck.Herant95} suggested that an $l=1$ mode might cause an
asymmetry that is sufficiently large to explain pulsar velocities even
in excess of 1000 km/s.

However, in previous simulations only anisotropies on relatively small
angular scales have been found. Consequently, the neutron star kicks
did not exceed 100-200 km/s (see e.g. \cite{scheck.JM96}). Motivated
by the hypothesis that the rapid onset of the explosion in these
models did not allow the merging of small-scale
anisotropies to global modes, we started to conduct a new study of
about 50 two-dimensional simulations, where the explosion, based on
the neutrino-heating mechanism, was triggered such that the initial
expansion set in more slowly \cite{scheck.Scheck+04}.

The neutron star core in these simulations was replaced by a boundary
condition. In contrast to earlier simulations, we prescribed more {\em
slowly} decaying and initially lower neutrino luminosities at this
boundary. This leads to longer explosion time scales and thus gives
convective structures between neutron star and shock more time to
merge. Shortly after the onset of the explosion most of our models are
dominated by an $l=1$ mode of the mass distribution: Infalling stellar
material is concentrated in one downflow and neutrino-heated matter
rises in one bubble on the opposite side. We found highly anisotropic
explosions for three different 15 solar mass progenitors. A rotating
progenitor (from \cite{scheck.Buras+03}) also developed low-mode
structures.

\begin{figure}
  \centerline{\epsfxsize=0.9\textwidth\epsffile{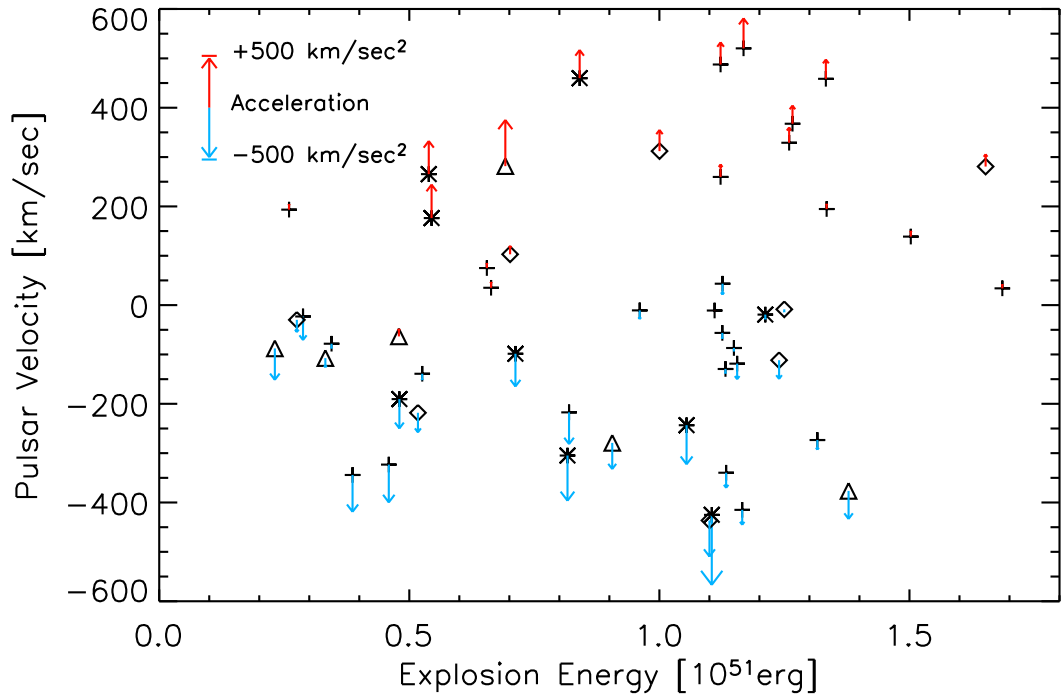}}
  \caption{Neutron star velocities and accelerations at one second
           after core bounce for a sample of simulations
           \cite{scheck.J04}. Different symbols denote different
           progenitor stars.}
  \label{scheck.vns-ans}
\end{figure}

Figure \ref{scheck.vns-ans} displays neutron star velocities and
accelerations one second after core bounce. We found neutron star
velocities as high as 520 km/s with still large acceleration after the
first second. The final neutron star velocities can therefore be
significantly higher. The dominant force that mediates the
acceleration is the gravitational attraction by the anisotropic
ejecta. Therefore the neutron star moves towards the most slowly
expanding ejecta, opposite to the main direction of the
explosion. Most of the acceleration takes place after the onset of the
explosion, substantial acceleration may continue for several seconds.

\begin{figure}
  \centerline{\epsfxsize=0.99\textwidth\epsffile{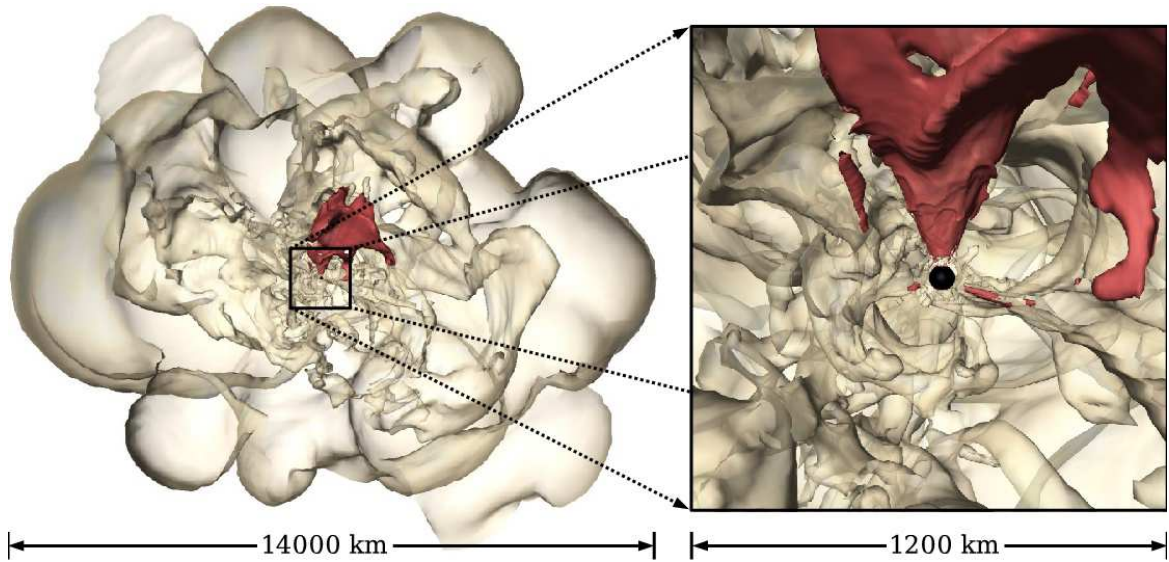}}
  \caption{Three-dimensional simulation \cite{scheck.thesis} one
second after core bounce. The bright structure is a surface of
constant proton-to-neutron ratio which roughly marks the outer
boundaries of the neutrino-heated high-entropy bubbles. The dark
surface, blown up in the right figure, is defined by a constant value
for the mass flux per unit area and defines a downflow of matter
towards the neutron star, the surface of which is indicated by the
black sphere (corresponding to a density of $10^{11} {\rm g/cm}^3$).}
  \label{scheck.3d}
\end{figure}

It has to be shown that the merging of modes in three dimensions is as
efficient as in two.  A first 3D simulation with still relatively low
resolution also developed an $l=1$ mode and is therefore promising (see
figure \ref{scheck.3d}).  Simulations with higher resolution are in
progress \cite{scheck.thesis}.

\vspace*{0.5cm}

{\small
\subsection*{Acknowledgements}
We are grateful to K.~Nomoto, A.~Heger, S.~Woosley, and
M.~Limongi for providing us with their progenitor data.
Supercomputer time at the John von Neumann Institute for
Computing in J\"ulich and the Rechenzentrum Garching is
acknowledged. This work was supported by the
Sonderforschungs\-be\-reich 375 ``Astroparticle Physics''
of the Deutsche Forschungsgemeinschaft.
}

\end{document}